\documentclass[reprint,superscriptaddress,amsmath, amssymb,aps,prd,notitlepage,longbibliography,floatfix,nofootinbib,twocolumn]{revtex4-1}

\usepackage{amsmath}
\allowdisplaybreaks[4]
\usepackage{cancel}
\usepackage{extarrows}
\usepackage{tensor}     
\usepackage{float}
\usepackage[caption = false]{subfig} 
\usepackage[final]{graphicx}   
\usepackage[
colorlinks=true,        
citecolor=blue,         
linkcolor=blue,         
urlcolor=blue           
]{hyperref}             
\usepackage{bm}         
\usepackage{xcolor}     
\usepackage{lipsum}
\usepackage{color}      
\usepackage[utf8]{inputenc} 
\usepackage[section]{placeins} 
\usepackage{appendix}
\usepackage{units}
\usepackage[capitalise]{cleveref}
\usepackage{units}
\newcommand{\nc}{\newcommand*}

\nc{\xbar}{\bar{x}}
\nc{\rhoeq}{\rho_{\mathrm{eq}}}
\nc{\zeq}{z_{\mathrm{eq}}}
\nc{\tla}{\tilde{\lambda}}
\nc{\bt}{\beta}
\nc{\dt}{\delta}
\nc{\Dt}{\Delta}
\nc{\vj}{\vec{j}}
\nc{\vl}{\vec{l}}
\nc{\hx}{\hat{x}}
\nc{\hy}{\hat{y}}
\nc{\bj}{\bm{j}}
\nc{\mJ}{\mathcal{J}}
\nc{\mP}{\mathcal{P}}
\nc{\Msun}{M_\odot}
\nc{\av}[1]{\langle #1 \rangle}
\nc{\eq}[1]{Eq.~\eqref{#1}}
\nc{\al}{\alpha}
\nc{\Xstar}{X_{\ast}}
\nc{\fpbh}{f_{\mathrm{pbh}}}
\nc{\vth}{\vec{\theta}}
\nc{\vla}{\vec{\lambda}}
\nc{\vd}{\vec{d}}
\nc{\Mmin}{M_{\mathrm{min}}}
\nc{\rmd}{\mathrm{d}}
\nc{\mmin}{{m_{\mathrm{min}}}}
\nc{\mmax}{{m_{\mathrm{max}}}}
\nc{\mR}{\mathcal{R}}
\nc{\tmR}{\tilde{\mathcal{R}}}
\nc{\s}{\sigma}
\nc{\ogw}{\Omega_{\mathrm{GW}}}
\nc{\addref}{[\textcolor{red}{add ref}] }
\nc{\Om}{\Omega}
\nc{\gm}{\gamma}
\nc{\Gm}{\Gamma}
\nc{\gpcyr}{\mathrm{Gpc}^{-3}\,\mathrm{yr}^{-1}}
\nc{\Eq}[1]{Eq.~\eqref{#1}}
\nc{\Fig}[1]{Fig.~\ref{#1}}
\nc{\Table}[1]{Table~\ref{#1}}
\nc{\lvc}{LIGO/Virgo} 
\nc{\Sec}[1]{Sec.~\ref{#1}}
\nc{\eg}{\textit{e.g.~}}

\nc{\sovast}{Soviet Ast.}

\newcommand\dif{\mathrm{d}}

\begin{document}

\title{Probing Ultralight Tensor Dark Matter with the Stochastic Gravitational-Wave Background from Advanced LIGO and Virgo's First Three Observing Runs}

\author{Rong-Zhen Guo}
\email{guorongzhen@ucas.ac.cn}
\affiliation{School of Fundamental Physics and Mathematical Sciences, Hangzhou Institute for Advanced Study, UCAS, Hangzhou 310024, China}
\affiliation{School of Physical Sciences, 
    University of Chinese Academy of Sciences, 
    No. 19A Yuquan Road, Beijing 100049, China}

\author{Yang Jiang}
\email{corresponding author: jiangyang@itp.ac.cn}
\affiliation{School of Physical Sciences, 
    University of Chinese Academy of Sciences, 
    No. 19A Yuquan Road, Beijing 100049, China}
\affiliation{CAS Key Laboratory of Theoretical Physics, 
    Institute of Theoretical Physics, Chinese Academy of Sciences,Beijing 100190, China}
\author{Qing-Guo Huang}
\email{corresponding author: huangqg@itp.ac.cn}
\affiliation{School of Fundamental Physics and Mathematical Sciences, Hangzhou Institute for Advanced Study, UCAS, Hangzhou 310024, China}
\affiliation{School of Physical Sciences, 
    University of Chinese Academy of Sciences, 
    No. 19A Yuquan Road, Beijing 100049, China}
\affiliation{CAS Key Laboratory of Theoretical Physics, 
    Institute of Theoretical Physics, Chinese Academy of Sciences,Beijing 100190, China}


\begin{abstract}
Ultralight bosons are attractive dark-matter candidates and appear in various scenarios beyond standard model. They can induce superradiant instabilities around spinning black holes (BHs), extracting the energy and angular momentum from BHs, and then dissipated through monochromatic gravitational radiation, which become promising sources of gravitational wave detectors. In this letter, we focus on massive tensor fields coupled to BHs and compute the stochastic gravitational wave backgrounds emitted by these sources. We then undertake a search for this background within the data from LIGO/Virgo O1$\sim$ O3 runs. Our analysis reveals no discernible evidence of such signals, allowing us to impose stringent limits on the mass range of tensor bosons. Specifically, we exclude the existence of tensor bosons with masses ranging from $4.0\times10^{-14}$ to  $2.0\times10^{-12}$ eV at $95\%$ confidence level.

\end{abstract}
\maketitle

\textit{Introduction.}
The detection of gravitational waves (GWs) from the merger of binary black holes (BHs) \cite{LIGOScientific:2018mvr} and neutron stars \cite{LIGOScientific:2020ibl} ushered in a new era of GW astronomy. Nowadays, GWs have been used to test our basic understanding of different aspects of fundamental physics \cite{Barack:2018yly,Baibhav:2019rsa,LIGOScientific:2019fpa,LIGOScientific:2020tif,Sathyaprakash:2019yqt,Jiang:2022mzt,Jiang:2022svq,Jiang:2022uxp}. Especially, it provides entirely new avenues for the detection and constraints of 
ultralight bosons\cite{Barack:2018yly,Baibhav:2019rsa,Yuan:2021ebu,Yuan:2022bem,Bertone:2019irm,Brito:2017wnc,Brito:2017zvb,Isi:2018pzk,Tsukada:2018mbp,Brito:2013wya}, such as spin-0 QCD axions, axion-like particles in string axiverse and spin-1 dark photons \cite{Arvanitaki:2009fg,Arvanitaki:2010sy,Weinberg:1977ma,Peccei:1977hh,Marsh:2015xka,Holdom:1985ag}, which are important candidates of dark matter (DM). 

The coupling of these fields to Standard Model particles is very weak, making direct detection in the laboratory difficult \cite{Marsh:2015xka,Feng:2010gw}. However, if the ultralight bosonic field with mass $m_b$ exists around a spinning BH, it can trigger superradiant instability if its typical oscillation frequency $\omega_R=m_b c^2/\hbar$ satisfies $0<\omega_R<m\Omega_H$, where $c$ is speed of light, $m$ is the azimuthal quantum number of the unstable mode and $\Omega_H$ is  horizon angular velocity of the BH (see \cite{Brito:2015oca} for a review). This mechanism results in the transfer of angular momentum and energy from the BH to the bosonic field. Ultimately, when $\omega_R\sim m \Omega_H$, superradiant instability suppresses and  corotating and non-axisymmetric boson condensate forms. The condensate is dissipated through the emission of almost monochromatic GWs with frequency $f_0\sim \omega_R/\pi$ \cite{Arvanitaki:2014wva,Brito:2014wla}. Thus, it leads to a wealth of  observable astrophysical phenomena \cite{Brito:2015oca,Dolan:2007mj,Brito:2013wya,Brito:2017zvb,East:2017ovw,Brito:2017wnc}, which have aroused much attention to probe ultralight bosons or the nature of BHs recently\cite{Guo:2023mel,Guo:2021xao,Jia:2023see,Zhou:2023sps,Cardoso:2018tly,Xia:2023zlf,Cannizzaro:2023jle}. 

Until now, there have been no positive results from direct detection of such nearly monochromatic GW signals \cite{Palomba:2019vxe,Tsukada:2020lgt,Ng:2020ruv,LIGOScientific:2021rnv,Sun:2019mqb}. Another effective method to constrain the parameters of ultralight bosons is to exploit the potential abundance of faint signals in the universe, that is, to search for stochastic GW background (SGWB) they generate. From an astronomical view, this background is generated by all the unresolved sources in Universe. During O1$\sim$ O3 observing runs of LIGO and Virgo, the upper intensity of SGWB has been set \cite{LIGOScientific:2016jlg,LIGOScientific:2019vic,KAGRA:2021kbb}. The SGWB  from BH-boson system was calculated by \cite{Brito:2017wnc,Brito:2017zvb}. So far, there have been studies utilizing SGWB to constrain the parameters of ultralight bosons \cite{Brito:2017wnc,Tsukada:2018mbp,Tsukada:2020lgt,Yuan:2021ebu,Yuan:2022bem}. The authors of \cite{Tsukada:2018mbp} and \cite{Tsukada:2020lgt}  provide the first search of SGWB in the first observing run of LIGO sourced by scalar and vector boson, respectively. However, the parameter constraints for ultralight tensor dark matter\cite{Aoki:2016zgp,Babichev:2016hir,Babichev:2016bxi,Aoki:2017cnz} have not been thoroughly investigated as extensively as scalar and vector dark matter \cite{Wu:2023dnp}.

In this letter, we explore the impact of ultralight tensor dark matter, which can be described by linearized ghost-free bimetric theory, and calculate the energy spectra of corresponding SGWB. Bayesian framework is adopted to search for such signal in LIGO and Virgo's observing period. From now on we use the units $G=c=\hbar=1$.
\medskip

\textit{Superradiant instability of ultralight tensor dark matter.} 
The dynamics of tensor dark matter can be described by spin-2 field $H_{\alpha\beta}$ on BH background with initial mass $M_i$ and initial dimensionless angular momentum $\chi_i=J_i / M_i^2$. This field can be regarded as linear massive tensor perturbation in the ghost-free bi-metric gravity theory \cite{Brito:2020lup,Babichev:2016bxi}, which equation of motion on Ricci-flat background reads:
\begin{equation}
\begin{aligned}
\square H_{\alpha \beta}+2 R_{\alpha \gamma \beta \delta} H^{\gamma \delta}-m_b^2 H_{\alpha \beta} & =0, \\
\nabla^{\alpha} H_{\alpha \beta}=0, \quad H^{\alpha}{ }_{\alpha} & =0,
\end{aligned}
\end{equation}
where the box operator, Riemann tensor $R_{\alpha \gamma \beta \delta}$, contractions are constructed with the background metric, and $m_b$ is the mass of the spin-2 field.

For ultralight ($m_b M\ll 1$) tensor dark matter, there are several different mechanisms for triggering instability. There exists unstable monopolar ($m=0$) modes, which are also existed in the spherical BH case. Such instabilities appear first, affecting the spacetime geometry of BHs and eventually forming hairy BHs \cite{East:2023nsk,Brito:2013wya,Babichev:2013una}. This instability has an extremely short characteristic timescale ($\tau_{\textrm{mono}}\sim m_b^{-1}$) \cite{Brito:2013wya}, and due to its monopolarity, no GW produces during this process. Therefore, the impact of this process can be ignored. And subsequent superradiant instability leads to the formation of the quasi-bound state, which characteristic frequency is complex \cite{Brito:2013wya,Dias:2023ynv}:
\begin{equation}
    \begin{aligned}
\frac{\omega_R}{m_b} & \simeq 1-\frac{(m_b M)^2}{2(\ell+n+S+1)^2}, \\
\omega_I & \propto (m_b M)^{4 \ell+5+2 S}\left(\omega_R-m \Omega_{\mathrm{H}}\right),
\end{aligned}
\end{equation}
where $\ell \geq 0, n \geq 0$, and the integer $S=(0, \pm 1, \pm 2)$ are the mode total angular momentum, overtone number, and polarization with $m \in[-\ell, \ell]$. We use the result on Kerr background, because these modes are significant support at $r\sim M (m_b M)^{-2}$ \cite{Brito:2013wya}, the deformation compared to the Kerr BH for the hairy BH is therefore negligible under ultralight limit. It is clearly that for $m\geq 1$ cases, if $\omega_R-m \Omega_H<0$, these states would grow exponentially until reaching the saturation point, $\omega_R \sim m \Omega_H$ in timescale $\tau_{\text {inst }} \equiv 1 / \omega_I$. In this letter, we concentrate on two dominant unstable appearing in non-relativistic approximation, a dipole mode ($l=m=1,S=0$) and a quadrapole mode ($l=m=2,S=-2$). Although they share the comparable timescale, it has been shown \cite{Ficarra:2018rfu} that the dipole mode can be reabsorbed by the BH, giving back almost all its mass and spin. The loss of mass and angular momentum is thus almost entirely attributable to the quadrapole mode itself. 

According to the conservation of mass and angular momentum, we can get the relationship between initial BH mass $M_i$, spin $J_i$ and the mass $M_f$ and spin $J_f$ when superradiant instability saturated at $\omega_R=m \Omega_H$:
\begin{equation}
J_f=J_i-\frac{m}{\omega_R}\left(M_i-M_f\right)=J_i-\frac{m}{\omega_R} M_T^{\mathrm{max}} ,
\end{equation}
\begin{equation}
M_f=\frac{m^3-\sqrt{m^6-16 m^2 \omega_R^2\left(m M_i-\omega_R J_i\right)^2}}{8 \omega_R^2\left(m M_i-\omega_R J_i\right)},
\end{equation}
where $M_T^\mathrm{max}$ is the maximal total mass of tensor dark matter. This quasi-bound state dissipates its energy by GWs in a typical timescale $\tau_{\mathrm{GW}}$:
\begin{equation}
\tau_{\mathrm{GW}}=M_f\left(\frac{d \tilde{E}}{d t} \frac{M_T^{\max }}{M_f}\right)^{-1},
\end{equation}
where $d \tilde{E}/d t$ is the reduced GW flux. In the qradrapole case, it reads \cite{Brito:2020lup}
\begin{equation}
    \tau_{\mathrm{GW}}\sim 290\left(\frac{M_f}{30\Msun}\right)^2 \left(\frac{3\Msun}{M_T^{\mathrm{max}}}\right)\left(\frac{m_b M_f}{0.2}\right)^{-10}\ \textrm{sec}.
\end{equation}
Since the hierarchy of timescales $\tau_{\mathrm{GW}} \gg \tau_{\text {inst }} \gg M$, we can get the total GW energy of the whole system emitted in a duration time $\Delta t=t_0-t_f$ by quasi-adiabatic approximation \cite{Brito:2017zvb}
\begin{equation}
    E_\text{GW}=\frac{M_T^\mathrm{max}\Delta t}{\Delta t + \tau_\mathrm{GW}},  \label{eq:E_GW}
\end{equation}
where $t_0$ is the age of the Universe and $t_f$ characterizes the time when the hairy $\mathrm{BH}$ is formed.

\medskip
\textit{SGWB from boson clouds.}
In this section, we aim to model the SGWB from the superradiant instability. As described in \Eq{eq:defomega} \cite{PhysRevD.59.102001}
\begin{equation}
    \Omega_\text{GW}(f)=\frac 1{\rho_c}\frac{\dif \rho_\text{GW}}{\dif \ln f},  \label{eq:defomega}
\end{equation}
it is characterized by the energy density per logarithm frequency and $\rho_c$ is the critical energy density required for a spatially flat Universe. To figure out this background, we need to sum the GWs radiated by all the possible sources. In our work, it is the superposition of all the individual BH-boson system triggering superradiance. The general formula is
\begin{equation}
    \Omega_\text{GW}(f)=\frac f{\rho_c}\int \mathrm{d}z\frac{\mathrm{d}t}{\mathrm{d}z}\int \mathrm{d}\bm{\theta}\; \mathcal{R}(\bm{\theta};z)\frac{\mathrm{d}E_s}{\mathrm{d}f_s}(\bm{\theta};m_b,f),
\end{equation}
where $\dif t/\dif z$ is the derivative of lookback time over redshift. It is determined by the $\Lambda\text{CDM}$ cosmology model \cite{Condon_2018}:
\begin{equation}
    \frac{\dif t}{\dif z}=\frac{1}{(1+z)H_0\sqrt{\Omega_M (1+z)^3+\Omega_\Lambda}}.  \label{eq:dtdz}
\end{equation}
$\mathcal{R}(\bm{\theta};z)$ is the formation rate of BH with parameter $\bm{\theta}$ per comoving volume at redshift $z$. $\dif E_s/\dif f_s$ is the energy spectrum of a single event in the source frame. Since the GW radiation is nearly monochromatic, the spectrum is well approximated by
\begin{equation}
    \frac{\dif E_s}{\dif f_s}=E_\text{GW}\delta(f_s-f_0)=E_\text{GW}\delta[f(1+z)-f_0],
\end{equation}
where $f_0=\omega_R/\pi$ and we replace the source-frame frequency $f_s$ by detected frequency due to the redshift of the sources.

Two kinds of BH populations are considered in our work, which are isolated stellar origin BHs (SBHs) and binary SBH merging remnants. For isolated BH channel, the formation rate is
\begin{equation}
    \mathcal{R}^{\text{iso}}(\chi, M; z)= p(\chi)R^\text{iso}(M;z).    \label{eq:mathcalR_iso}
\end{equation}
By writing down \Eq{eq:mathcalR_iso}, we have separate the relation of BH spin since its distribution can not be derived from theory and here we adopt a uniform prior distribution:
 \begin{equation}
     p(\chi) = \begin{cases}
     \frac1{\chi_\mathrm{u}-\chi_\mathrm{l}},& \chi_\mathrm{l}\leq\chi\leq\chi_\mathrm{u} \\
     0, & \text{otherwise.}
     \end{cases}
 \end{equation}
 The remaining part can be determined by astrophysical model
 \begin{equation}
     R^\text{iso}(M;z) = \psi(z_f)\frac{\xi(M_*)}{M_*}\frac{\dif M_*}{\dif M}, \label{eq:R_iso}
 \end{equation}
where $M_*$ is the mass of progenitor star. Its population is the product of star formation rate and initial mass function. We
adopt the Salpeter initial mass function \cite{1955ApJ...121..161S}, which is a power law $\xi(M_*)\propto M^{-1.35}$ and normalized between $0.1\Msun\sim100\Msun$. For star formation rate, we use a model proposed in \cite{2003MNRAS.339..312S}
\begin{equation}
    \psi(z)=\nu\frac{a\exp[b(z-z_m)]}{a-b+b\exp[a(z-z_m)]}.
\end{equation}
As suggested in \cite{Vangioni:2014axa}, the fitting parameters are taken to be $\nu=0.178\,\Msun/\text{yr}/\text{Mpc}^3$, $a=2.37$, $b=1.80$ and $z_m=2.00$. $z_f$ is the redshift at progenitor birth, so the corresponding lookback time satisfies $t_f=t-\tau(M_*)$, where $\tau(M_*)$ denotes the lifetime of star and we calculate this quantity based on \cite{Schaerer:2001jc}. Besides, to determine the derivative item in \Eq{eq:R_iso}, we use the relation of mass between BH and progenitor star metalliticity in \cite{2012ApJ...749...91F,Belczynski:2016obo}.

For merging remnant channel, 
\begin{equation}
    \mathcal{R}^\text{rem}(m_1, m_2; z)=p(m_1, m_2)R^\text{rem}(m_1,m_2;z).   \label{eq:R_rem}
\end{equation}
The adopted mass distribution of binary system is a broken power law model \cite{LIGOScientific:2020kqk,KAGRA:2021kbb}. That is
\begin{gather}
    p(m_1, m_2) \propto\begin{cases}
    m_1^{-1.58}, & m_\text{mim}\leq m_1<m_*\\
    m_1^{-5.59}, & m_*\leq m_1<m_\text{max}\\
    0, & \text{otherwise}\end{cases}\notag \\
    m_\text{min}=3.96\Msun,\notag\\
    m_\text{max}=87.14\Msun,\notag \\
    m_*=m_\text{min}+0.43(m_\text{max}-m_\text{min}).\notag
\end{gather}
$R^\text{rem}$ appeared in \Eq{eq:R_rem} is the binary merger rate. It is the convolution of the binary formation rate with the distribution of time delays between formation and merger:
\begin{equation}
    R^\text{rem}(m_1,m_2;z)=\int_{t_\text{min}}^{t_\text{max}} R_f(m_1, m_2;z_f)p(t_d)\,\dif t_d.
\end{equation}
Following \cite{LIGOScientific:2016fpe,LIGOScientific:2017zlf,KAGRA:2021kbb}, we have assumed that $R_f(m_1,m_2;z_f)$ is proportional to a metallicity-weighted star formation rate. The parameters of weighting rectification are the same as \cite{KAGRA:2021kbb} and the time distribution is taken as $p(t_d)\propto t_d^{-1}$ with $50\,\text{Myr}\leq t_d\leq13.5\,\text{Gyr}$. Numerical Relativity suggests that the mass and spin of BH remnant are \cite{Berti:2007fi,Scheel:2008rj,Barausse:2009uz}
\begin{align}\begin{split}
    M &= m_1+m_2-(m_1+m_2)\Bigg[\left(1-\sqrt{\frac89}\right)\nu \\
    &\phantom{=}-4\nu^2\left(0.19308+\sqrt{\frac89}-1\right)\Bigg],
    \end{split}\notag \\
    \chi &= \nu\left(2\sqrt{3}-3.5171\nu+2.5763\nu^2\right).
\end{align}
The local merger rate is normalized as \cite{KAGRA:2021kbb}
\begin{equation}
    \int \mathcal{R}^\text{rem}(\bm{m};0)\,\dif\bm{m} = 19\;\text{Gpc}^{-3}\text{yr}^{-1}.
\end{equation}
\begin{figure}[ht]
    \centering
    \includegraphics[width=0.9\columnwidth]{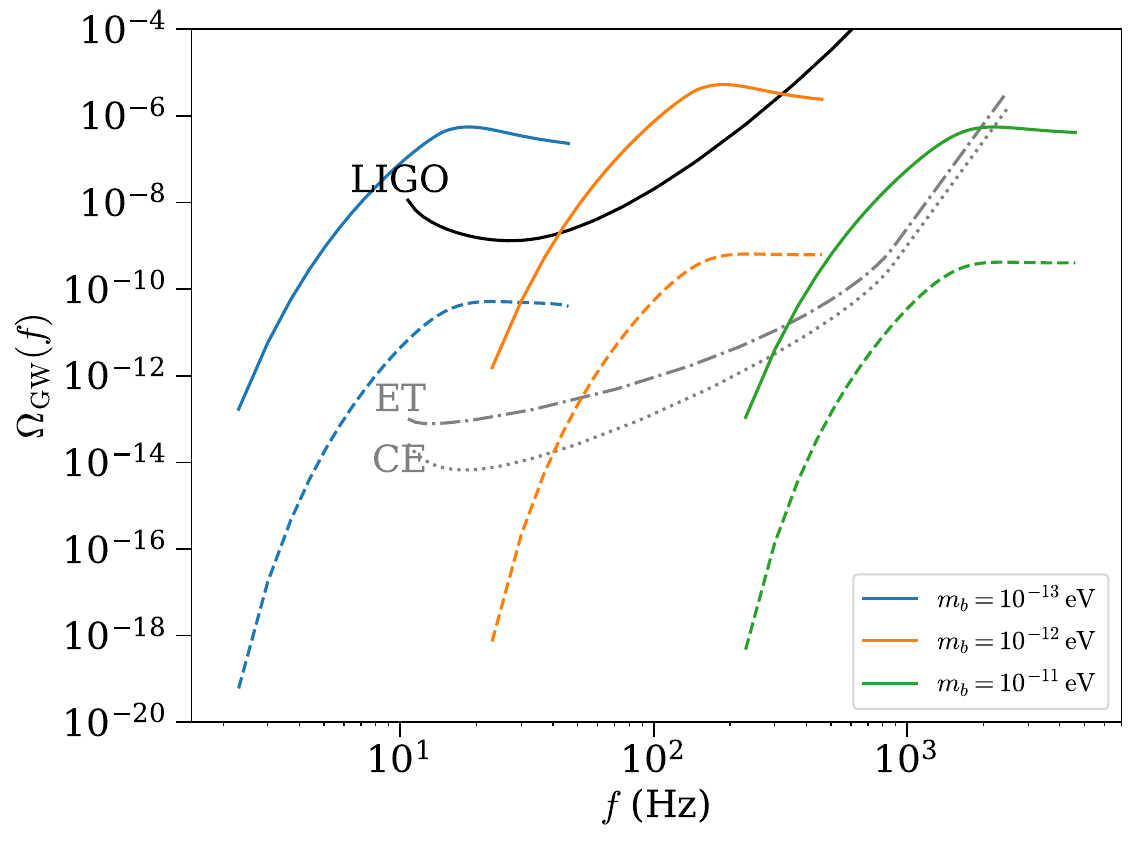}
    \caption{The energy spectra of SGWB from superradiant instability of spin-2 field. Solid curves are backgrounds from isolated BH channel with uniform prior $\chi\in[0,1]$ while dashed curves represent BBH remnant contributions. The power-law integrated sensitivity curves \cite{Thrane:2013oya} of LIGO O3 run, Einstein Telescope (ET) \cite{LIGOScientific:2016wof} and Cosmic Explorer (CE) \cite{Punturo_2010} are also shown.}
    \label{fig:spectrum}
\end{figure}

Add the contributions from these two kinds of BH populations up and we work out the actual background. Typical spectra within LIGO's frequency band are shown in \Fig{fig:spectrum}. We show the spectra of these two channels separately. The background is dominated by isolated BH population. Its amplitude is about 4 orders of magnitude larger than BBH remnant channel.

\medskip
\textit{Data analysis and results.}
SGWB will cause a correlation in the output of detector pairs. An estimator can be constructed with Fourier transform of the strain data as follows \cite{Romano:2016dpx}
\begin{equation}
    \hat{C}_{IJ}(f)=\frac{2}{T}\frac{\text{Re}[\tilde{s}_I^*(f)\tilde{s}_J(f)]}{\gamma_{IJ}(f)S_0(f)}, \label{eq:C_IJ}
\end{equation}
where $S_0(f)=(3H_0^2)/(10\pi^2f^3)$. $\gamma_{IJ}(f)$ is the overlap reduction function \cite{PhysRevD.59.102001} of detector pair $IJ$ and $T$ denotes observing time. \Eq{eq:C_IJ} has been normalized to $\langle \hat{C}_IJ(f) \rangle=\Omega_\text{GW}(f)$ and the variance is
\begin{equation}
    \sigma_{IJ}^2(f) = \frac{1}{2T\Delta f}\frac{P_I(f)P_J(f)}{\gamma_{IJ}^2(f)S_0^2(f)}.   \label{eq:sigma}
\end{equation}
$P_{I,J}(f)$ appeared in \Eq{eq:sigma} are the power spectral densities of the interferometers. To evaluate the mass of spin-2 boson, we perform Bayesian analysis. After glitch gating and non-stationary rejection, the noise of detectors is well approximated by uncorrelated Gaussian signal. Therefore, the total likelihood is 
\begin{equation}
     p(\bm{\hat{C}}|\bm{\theta})\propto \exp\left[-\sum_{IJ}\sum_{f}\frac{\left(\hat{C}_{IJ}(f)-\Omega(f;\bm{\theta})\right)^2}{2\sigma^2_{IJ}(f)}\right],
\end{equation}
where $\bm{\theta}$ denotes a sequence of parameters to be determined by analysis and we have multiplied the likelihood from all the frequency bins and detector pairs. The ratio of model evidence
\begin{equation}
    \mathcal{B}_\text{\,noise}^\text{\,model}=\frac{p(\bm{\hat{C}}\,|\,\text{Model of signal})}{p(\bm{\hat{C}}\,|\,\text{Pure noise})},
\end{equation}
so-called Bayes factor illustrates the statistical significance of the presence of the signal.
\begin{figure}[ht]
    \centering
    \includegraphics[width=0.9\columnwidth]{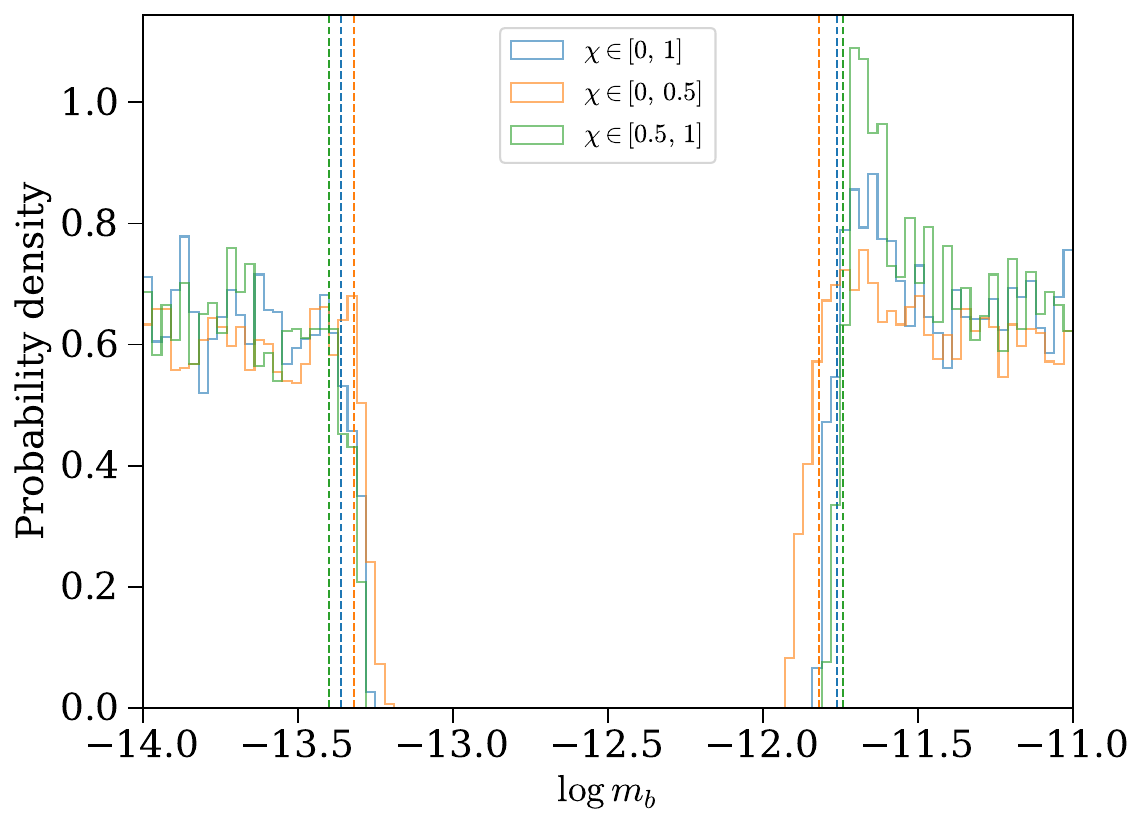}
    \caption{Posterior distributions of $\log m_b$ assuming a uniform spin distribution for the isolated SBHs. The orange dash lines denote $95\%$ exclusion intervals.}
    \label{fig:hist}
\end{figure}

We set a log-uniform prior between $10^{-14}\sim10^{-11}\,\text{eV}$ to the mass of tensor field. To figure out the influence of spin distribution, we have assumed three different combinations of $\chi_\mathrm{l,u}$ and treated their energy spectrum separately. As a result, the logarithm of Bayes factors are $-0.29\sim-0.26$, indicating that there are no such signals in the observing periods. The probability distributions of the mass and $95\%$ exclusion ranges by assuming a uniform distribution of spin between $[0,\,1],\;[0,\,0.5]$ and $[0.5,\,1]$ are displayed in \Fig{fig:hist}. The corresponding excluded range of $\log (m_b/\text{eV})$ is around $-13.4\sim-11.7$ and it does not a show strong connection with the spin assumptions we adopted. Compared to scalar and vector fields, the observing data puts a stronger constraint on the mass of tensor field. This is within our expectations as tensor field  has a shorter timescale of superradiant instability comparing with scalar and vector fields with same mass \cite{Brito:2015oca,East:2017mrj}. 

\medskip

\textit{Discussions and conclusions. }
In this work, we search for the SGWB signal produced by ultralight tensor dark matter around both isolated SBHs and  binary SBHs remnants in LIGO/Virgo O1$\sim$ O3 runs. We find no evidence to suggest the presence of the signal then we place constraints on the mass of tensor bosons. Assuming a uniform distribution of spin between $[0,\,1],\;[0,\,0.5]$ and $[0.5,\,1]$ for isolated SBHs, we exclude the mass range $10^{-13.4}\sim10^{-11.7}\text{eV}$. The relationship between the excluded mass range and the spin distribution is not very tightly connected.

It is noteworthy that we notice recent advances in the purely numerical spectral analysis of ultralight tensor dark matter \cite{East:2023nsk,Dias:2023ynv}. These studies contain several new modes beyond the hydrogenic approximation we used in the work. Since these modes stem from nonseparability of the field equations, their dynamical impact becomes very difficult to compute. This aspect remains a subject for future investigations and warrants careful consideration in subsequent research endeavors.

\textit{Acknowledgements.}
 We thank Yuan Chen for useful discussions. QGH is supported by the grants from NSFC (Grant No.~12250010, 11975019, 11991052, 12047503), Key Research Program of Frontier Sciences, CAS, Grant No.~ZDBS-LY-7009, the Key Research Program of the Chinese Academy of Sciences (Grant No.~XDPB15). We acknowledge the use of HPC Cluster of ITP-CAS.

\bibliography{refs}
\end{document}